\documentstyle[aps,epsfig,prl]{revtex}

\begin{document}

\title{A new test for random number generators:\\
Schwinger-Dyson equations for the Ising model}

\author{
H.~G.~Ballesteros and
V.~Mart\'{\i}n-Mayor,\\
{\it Departamento de F\'{\i}sica Te\'orica I,}\\ 
{\it Universidad Complutense de Madrid,}\\
{\it 28040 Madrid, Spain,}\\
{\tt\small  hector, victor@lattice.fis.ucm.es}
}
\date{\today}

\maketitle

\begin{abstract}
We use a set of Schwinger-Dyson equations for the Ising Model to check
several random number generators. For the model in two and three
dimensions, it is shown that the equations are sensitive tests of bias
originated by the random numbers. The method is almost costless in
computer time when added to any simulation.
\end{abstract}

\bigskip


\section{Introduction}

Among the sources of systematic error in Monte Carlo (MC) simulations,
the most frightening is the lack of randomness in the used pseudo
random number generator (PRNG). Indeed, in a modern MC simulation as
much as $10^{13}$ random numbers may be generated~\cite{PERC}. This is
as long as the longest test of random numbers we have heard
of~\cite{VATTU}. Therefore, a PRNG needs to be fast and thus not too
sophisticated, but it also should not bias the simulation
results. Shift-register PRNG's~\cite{R250} have become very popular,
due to their speed, but they have been shown to be unreliable for some
applications~\cite{PARISIRAPUANO}.  The study of the trustworthiness
of a PRNG is quite difficult as the answer is problem-dependent,
algorithm-dependent and (most important) precision-dependent. For
instance, in ref.~\cite{DPLANDAU} some commonly used shift-register
PRNG's were shown to yield incorrect results for the two dimensional
Ising model simulated with the Wolff's single-cluster
algorithm~\cite{WOLFF}. Of course, this failure is related to one's
statistical accuracy (all the generators in ref.~\cite{DPLANDAU} would
be ``correct'' with $5\%$ errors).  In particular, the {\bf R250}
shift-register PRNG was found to be very dangerous for single-cluster
update, but safe for its use with the Metropolis algorithm. Not long
after that, {\bf R250} was shown to fail in the Metropolis update of
the Blume-Capel model for some lattice
sizes~\cite{BLUMECAPEL}. Another example of the difficulty in
certifying PRNG's can be found in ref.~\cite{CODDINGTON}. There, {\bf
RANF} (the standard Cray PRNG) is shown to be ``very good'' in the
author own wording. This means that the longest carried run did not
find bias in a two dimensional Ising model simulation, where
comparison with the exact solution is
possible~\cite{FISHER}. Nevertheless, it has produced awfully wrong
results in a U(1) lattice gauge-theory simulation~\cite{KOGUT}.
Moreover, it is fairly common that one's simulation is, itself, the
longest run ever carried for this particular problem (otherwise, why
bother doing it?).  Unless independent, algorithmically different
simulations were performed, it is clear that one's result will not be
yet {\it established}. Further confidence can be obtained if sensible
consistency tests are also carried. In this paper we want to show
that Schwinger-Dyson identities may be useful in this respect,
specially when no exact solution is at hand.  Let us finally mention
that the investigation of the reasons for PRNG induced bias is
interesting in itself~\cite{BLOTE}, but it has not yet reached
predictive power (one wants to know {\it before} carrying the
simulation).

\section{The Equations}

Generally speaking, Schwinger-Dyson equations are relations of
the type
\begin{equation}
0=\left\langle\frac{\delta O}{\delta \phi (x)}\right\rangle-
\left\langle O \frac{\delta H}{\delta \phi (x)}\right\rangle,
\label{SDCONT}
\end{equation}
where $O$ is an arbitrary operator and $H$ is the hamiltonian 
(notice however that for Eq. (\ref{SDCONT}) not to be a trivial $0=0$
statement, $O$ should be an odd operator if $H$ is symmetric under the
$\phi\rightarrow -\phi$ transformation).
The problem is that the longest MC
runs are usually done in  discrete spin models,
for which there are no continuous variables. Nevertheless, for spin
models the measure usually has a Z$_2$ symmetry, which allows
to obtain equations analogous to (\ref{SDCONT}).
As an example, let us consider the Ising model on the
cubic lattice, with nearest neighbors interaction.
The Hamiltonian is
\begin{equation}
H=-\beta\sum_{<i,j>} \sigma_i \sigma_j\ ,
\label{HAMILTONIANO}
\end{equation}
where $\sigma$ are the usual Z$_2$ spin variables. Let us 
call $S_i$ to the sum of the spins coupled to spin $\sigma_i$. 
The self-evident relation
$$\sum_{\sigma = -1, 1} f(\sigma) = \sum_{\sigma = -1, 1} f(-\sigma),$$
yields for any observable 
depending on the spin $\sigma_i$ (and possibly also on others), 
$O(\sigma_i;...)$, the following relation
\begin{equation}
\left\langle O(\sigma_i;\cdots)\right\rangle=
\left\langle O(-\sigma_i;\cdots){\mathrm e}^{-2\beta\sigma_i S_i}
\right\rangle.
\label{SD}
\end{equation}
In particular, one gets
\begin{eqnarray}
1&=&\left\langle {\mathrm e}^{-2\beta\sigma_i S_i}\right\rangle\, , 
\label{UNO}\\
\left\langle \sigma_i\sigma_j\right\rangle&=&
-\left\langle \sigma_i\sigma_j{\mathrm e}^{-2\beta\sigma_i S_i}
\right\rangle\ +
\ 2\,\delta_{ij},\label{SDCORR}
\end{eqnarray} 
where $\delta_{ij}$ is the Kronecker symbol. In order to gain
statistics, it is useful to sum Eq. (\ref{UNO}) for all the lattice
sites (the lattice size being $L$, its volume is $V=L^D$). One
obtains:
\begin{equation}
1=\left\langle\frac{1}{V}\sum_i {\mathrm e}^{-2\beta\sigma_i
S_i}\right\rangle.
\label{UNOGLOBAL}
\end{equation}

Summing to the nearest neighbors in Eq.~(\ref{SDCORR}), we  obtain 
an expression which has been very useful in
MC Renormalization Group investigations of the dynamics of the
Poliakov loop in lattice gauge-theories~\cite{TONY}:
\begin{equation}
0=\left\langle\frac{1}{V}\sum_i \sigma_i S_i\,
(1+{\mathrm e}^{-2\beta\sigma_i S_i})\right\rangle.
\label{ENERGIA}
\end{equation}
It is trivial to generalize Eq.~(\ref{ENERGIA}) when more couplings
are included in the Hamiltonian, as needed in a MC Renormalization
Group study.

In addition, a non-local identity is obtained from
Eq. (\ref{SDCORR}) summing to all $i$ and $j$: 
\begin{equation}
0=-\frac{2}{V}\ +\ \left\langle\frac{1}{V^2}\sum_{i,j} \sigma_i 
\sigma_j\, (1+{\mathrm e}^{-2\beta\sigma_i S_i})\right\rangle.
\label{SDMAG}
\end{equation}

At this point, it is natural to ask if the right-hand side of
Eqs.~(\ref{UNOGLOBAL},\ref{ENERGIA},\ref{SDMAG}) can be measured
with reasonable statistical accuracy. We shall see that the answer
is positive. Then the next natural question to ask is if a PRNG inducing 
bias also spoils the fulfillment of these equations. We shall
find a positive answer only for Eqs. (\ref{UNOGLOBAL}) and (\ref{ENERGIA}). 

Finally, let us mention that the Z$_2$ symmetry is embedded in the
symmetry of many other models, therefore
Eqs.(\ref{UNOGLOBAL},\ref{ENERGIA},\ref{SDMAG}) hold as they are for
O($N$) spin-models, or, with trivial modifications, for SU($2N$)
lattice gauge-theories.

\section{Numerical results}

We have studied the Ising model (with periodic boundary conditions) in
two and three dimensions at their critical points. Three update
methods have been considered: Metropolis \cite{METROPOLIS}, the
Swendsen-Wang cluster method \cite{CLUSTER} and Wolff's
single-cluster (SC) \cite{WOLFF}. For each update, we have
employed three PRNG.  One has been the problematic~\cite{DPLANDAU}
{\bf R250}
\begin{equation}
X^{\mathrm {R250}}_n= (X^{\mathrm {R250}}_{n-103}+
X^{\mathrm {R250}}_{n-250})\ {{\mathrm {mod}}\  2^{32}}.
\label{PRNG250}
\end{equation}
The second has been the Parisi-Rapuano
({\bf PR}) PRNG~\cite{PARISIRAPUANO}, 
which has been found not  quite correct in four dimensional 
site-percolation~\cite{PERC}:
\begin{equation}
X^{\mathrm{PR}}_n=Y_n\  {\mathrm {xor}}\ Y_{n-61},
\label{PR}
\end{equation}
where
$$Y_n= (Y_{n-24}+Y_{n-55})\ {{\mathrm {mod}}\  2^{32}}.$$

Our last generator is defined with the help of a congruential
generator:
$$Z_{n+1}= (16807 \, Z_n ) \, {{\mathrm {mod}}\  (2^{31} -1)}.$$
Then, the {\bf PRC} PRNG \cite{PERC} is defined as
\begin{equation}
X^{\mathrm{PRC}}_n = (X^{\mathrm{PR}}_n\, +\, 2\, Z_n)\ {{\mathrm
{mod}}\  2^{32}}.
\label{PRC}
\end{equation}
Our statistics have been the following. In two dimensions we have
considered a $L=16$ lattice. We have simulated at the exact critical
point up to 6 digits
$$\beta_{\mathrm c}=0.440687.$$ We have measured every 20 Metropolis
sweeps or 20 single-clusters, performing $8\times 10^7$ Metropolis
full-lattice sweeps, and updating $4\times 10^7$ clusters. For the
Swendsen-Wang algorithm, we measure every 5 sweeps, and have generated
the clusters $4\times 10^7$ times.

In three dimensions, the critical coupling is known with great
accuracy \cite{TALAPOV}. We have simulated at
$$\beta=0.221654.$$ As shown in ref.~\cite{BLUMECAPEL}, it might
happen that the bias only appears for some lattice sizes. Therefore,
we have studied $L=16$ and $24$ lattices. For Metropolis or
single-cluster, we measure every 10 sweeps. We perform $10^7$
Metropolis sweeps, and generate $10^7$ clusters. In the Swendsen-Wang
case, we measure every 4 sweeps, generating the clusters $4\times
10^6$ times.  We have found quite clear results for the different
simulations, except for the single-cluster update of the $16^2$ and
$16^3$ lattices, with {\bf PR} as PRNG. We have found convenient to
extend these two simulations, although this is in principle a
dangerous procedure. Of course, one cannot proceed the run until the
results ``looke nice'', since this would bias the results. To avoid
subjective decisions, we have fixed {\em a priori} the total
(much longer than the initial)
simulation time: these two simulations have been 40 times
longer than the others. In this way, error bars shrink enough to
distinguish between a large statistical fluctuation and a systematic
error.

Before presenting our results, a word of caution is in order.  We have
carried 27 independent simulations (3 lattices $\times$ 3 PRNG
$\times$ 3 updates), so, the number of expected data points which are
more than one standard deviation away is uncomfortably
large. Specifically, one can easily estimate that the number of points
that are more than 1.7 deviations away ($10\%$ probability) must be between 2
and 4. Moreover, errors are not obtained with perfect
accuracy. Allowing a $10\%$ error in the error determination, we have
considered deviations larger than 3.3 error bars as a significant
signal of bias (less than a $0.13\%$ probability).

\begin{figure}[p]
\begin{center}
\leavevmode \epsfig{file=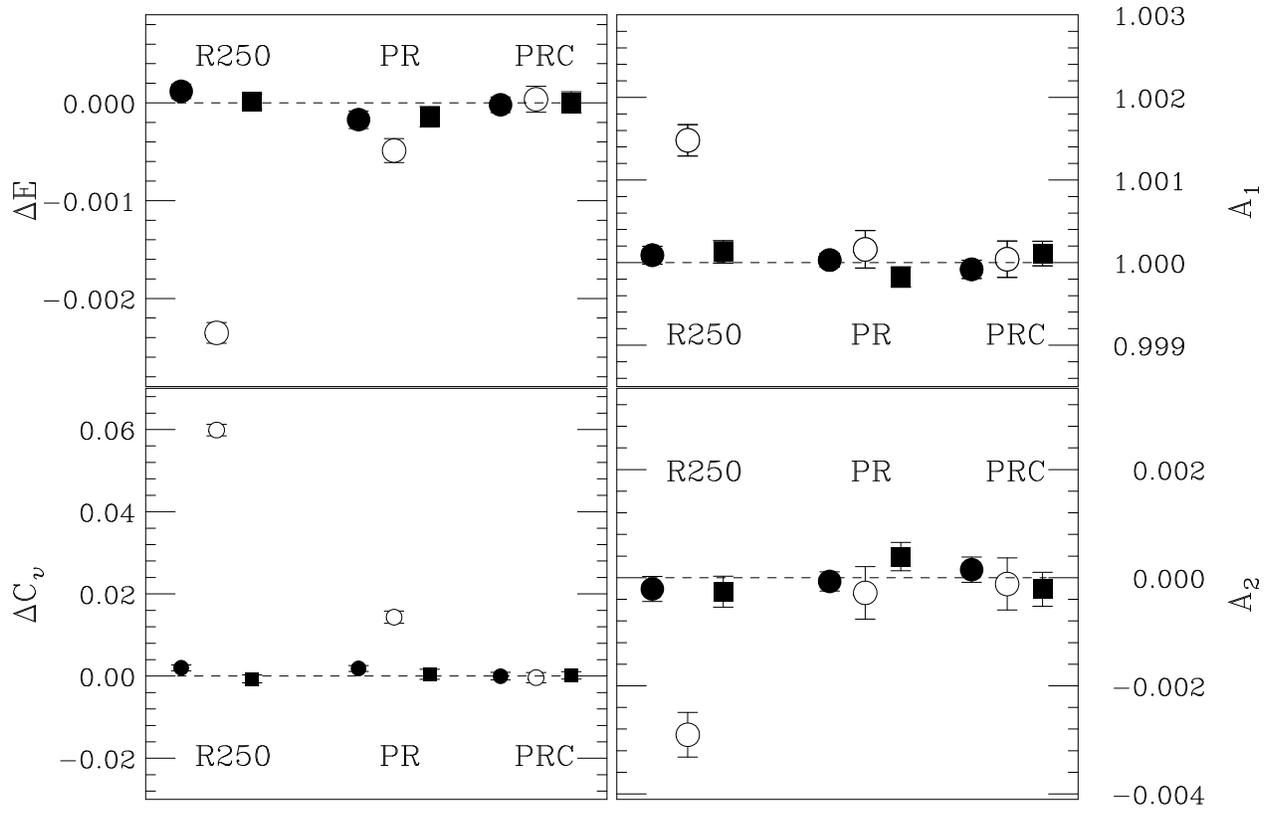,width=0.6\linewidth,angle=90}
\end{center}
\caption{Difference with the exact results 
for the energy and the specific-heat in
a $16^2$ lattice. We also plot  $A_1$ and $A_2$. 
Full circles correspond to Swendsen-Wang update, open ones to 
single-cluster and squares are from the Metropolis update.}
\label{162D}
\end{figure}

Let us first discuss our results in two dimensions.  In the left-hand
side of figure \ref{162D} we plot the deviations of the energy and the
specific-heat from their exact values \cite{FISHER}.  We find
significant deviations only for the single-cluster update when using
{\bf R250} and {\bf PR} as PRNG's (the former is not surprising
\cite{DPLANDAU}). It is clear that the exact solution is the best of
possible tests, but we would like to confront it with the
Schwinger-Dyson test. For this, let us define the quantities:
\begin{eqnarray}
{\mathrm A}_1&=&\left\langle\frac{1}{V}\sum_i {\mathrm e}
^{-2\beta\sigma_i S_i}\right\rangle_{\mathrm{MC}} \, ,\\\nonumber
{\mathrm A}_2&=&\left\langle\frac{1}{V}\sum_i \sigma_i S_i\,
(1+{\mathrm e}^{-2\beta\sigma_i S_i})\right\rangle_{\mathrm{MC}} \, ,
\label{E123}
\end{eqnarray}
which are the right-hand side of
Eqs.(\ref{UNOGLOBAL},\ref{ENERGIA}). Unfortunately Eq. (\ref{SDMAG})
has been found to hold within errors in all cases. In the above
expressions $\langle\ \rangle_{\mathrm{MC}}$ is the MC average, not
the expectation value.  We show our results for A$_1$ and A$_2$ in the
right-side of figure \ref{162D}.  The only significant deviation
found is in the single-cluster update with {\bf R250}.  This does not
mean that the Schwinger-Dyson identities can be fulfilled with a
biasing PRNG, as this is of course a matter of accuracy. In fact,
performing a 40 times longer run with {\bf PR}, we find
\begin{eqnarray*}
{\mathrm A}_1&=&1.00024(4) ,\\
{\mathrm A}_2&=&-0.00047(8) .\\
\end{eqnarray*}
Thus, both the exact solution test and the Schwinger-Dyson identities
test are failed by this SC-{\bf PR} combination, but the exact
solution test is more sensitive in this case.

We can discuss our results more quantitatively.  For small bias, it is
natural to expect that its main effect can be described as a shift on
the coupling, from $\beta$ to $\beta'=\beta-\Delta\beta$. With this
assumption, we can relate the different bias.  Let $\Delta O$ be the
the difference between the mean
value of $O$ obtained with some MC simulation, and its true Boltzmann
average, we obtain to first order in $\Delta \beta$
{\footnote{To avoid confusions let us recall that in this 
expression $E=\left\langle\frac{1}{2V}\sum_i \sigma_i S_i\right\rangle$, and it is a growing
function of $\beta$.}}
\begin{equation}
\Delta O \approx -\frac{\partial_\beta \langle O\rangle}{4 E}\Delta {\mathrm A}_1\, .
\label{SUPERTEST}
\end{equation}
In this way, we can understand that the bias for the energy has
opposite sign that the one for ${\mathrm A}_1$ (see footnote), and it
is also opposite to the bias for the specific heat (it is well known
that the maximum of the specific heat of the two dimensional Ising
model in a finite lattice is at $\beta<\beta_{\mathrm c}$).  The only
evidence that we can offer for Eq.~(\ref{SUPERTEST}) is empirical, and
it is shown in table \ref{SESGO2D}. Nevertheless, we find the
agreement quite satisfactory for such a rough calculation. Moreover,
from table \ref{SESGO2D} we can estimate that
\begin{equation}
{\begin{array}{cccc}
\Delta E &\approx & -1.6 &\Delta {\mathrm A}_1\, ,\\
\Delta C_v &\approx & 40 &\Delta {\mathrm A}_1\, ,
\end{array}}
\end{equation}
where the coefficient for the energy is really $1.58(19)$, to
be compared with 1.33 from Eq.~(\ref{SUPERTEST}). Notice that
if Eq.~(\ref{SUPERTEST}) could be rigourously established, it
would be enough to estimate the failure in the Schwinger-Dyson
test for any PRNG, to get the safe accuracy level for every
observable. However, to our knowledge, such an interesting property has
not been proved for any PRNG test.

\begin{figure}[p]
\begin{center}
\leavevmode \epsfig{file=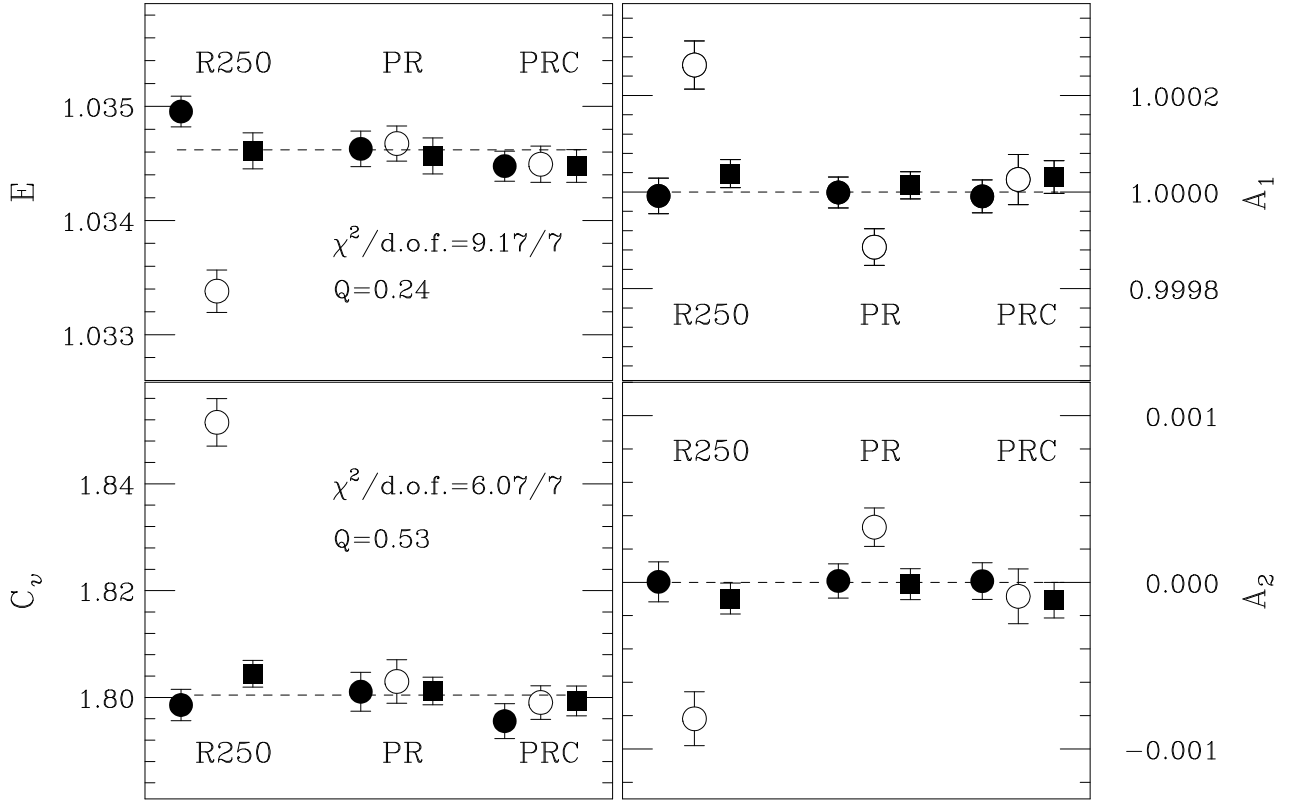,width=0.6\linewidth,angle=90}
\end{center}
\caption{Simulation results for the energy, the specific-heat, $A_1$
and $A_2$ in a $16^3$ lattice.  Dashed-lines for $E$ and $C_v$ are
obtained from a $\chi^2$ minimization, excluding the SC-{\bf R250}
data. Q is the probability of getting a larger value of $\chi^2$.
Full circles correspond to Swendsen-Wang update, open ones to
single-cluster and squares are from the Metropolis update.}
\label{163D}
\end{figure}

\begin{figure}[p]
\begin{center}
\leavevmode \epsfig{file=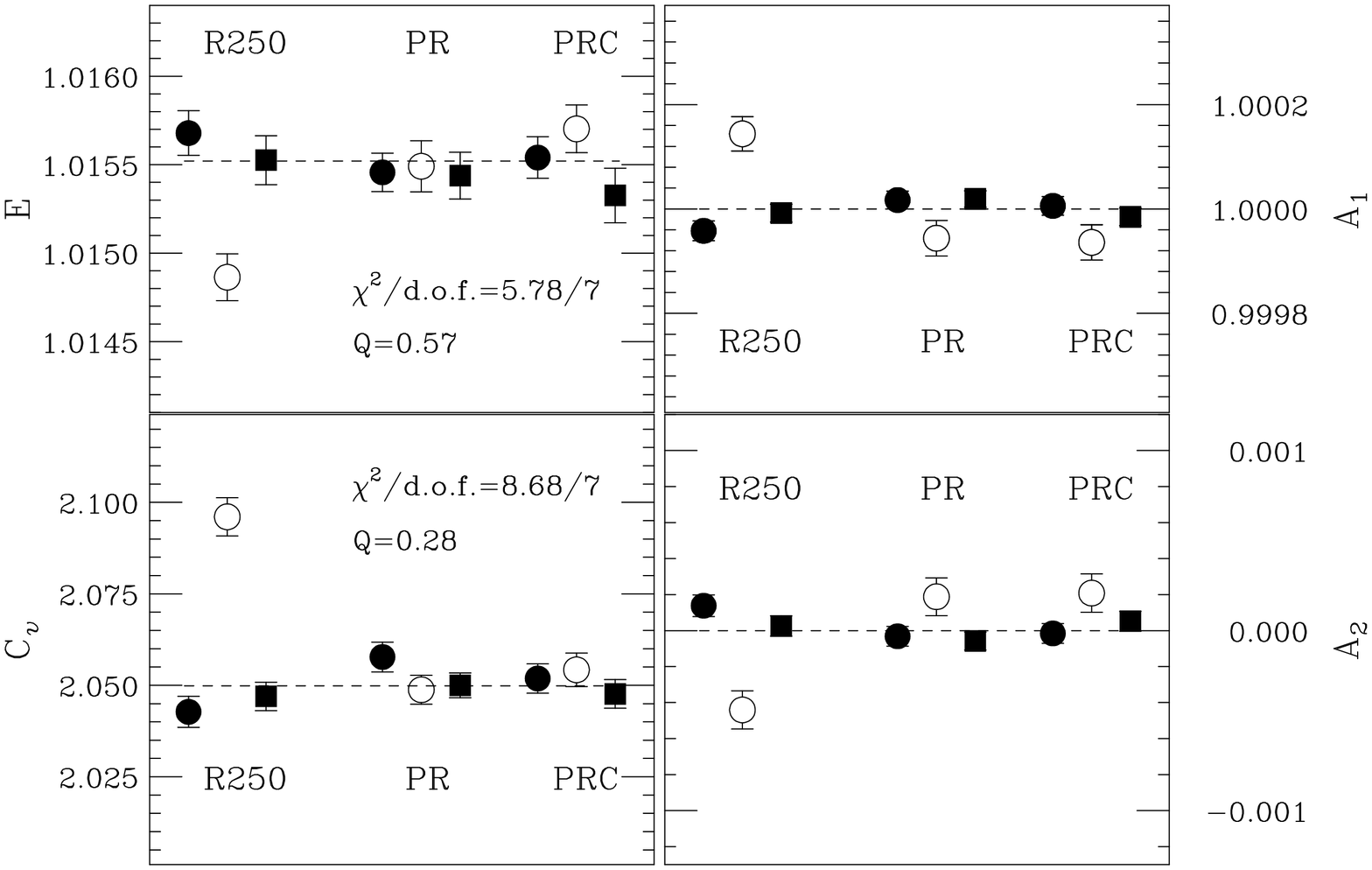,width=0.6\linewidth,angle=90}
\end{center}
\caption{Same as figure \ref{163D} for a $24^3$ lattice.}
\label{243D}
\end{figure}

For the three dimensional case, we plot our results for the energy and
the specific-heat in the left-hand side of figures \ref{163D} and
\ref{243D}.  In this case, we unfortunately lack an exact solution to
control for the bias.  However, we can study the statistical
compatibility of our data.  From the plot it is apparent that the
SC-{\bf R250} results are biased.  In the figures we show the data
with a weighted estimate of the energy and the specific-heat
(excluding the SC-{\bf R250} data). In fact, no further significant
deviations are found.  For the Schwinger-Dyson test, we find again
strong signal of bias in A$_1$ and A$_2$ for the combination of {\bf
R250} with single-cluster. We also find worrying deviations in the
single-cluster update with {\bf PR} as PRNG for $L=16$. To clarify if
a bias is present in this case, we have performed a 40 times longer
run. The new results are
\begin{eqnarray*}
{\mathrm A}_1&=& 0.999966(8),\\ {\mathrm A}_2&=& 0.000101(26).
\end{eqnarray*}
Thus, the {\bf PR} PRNG does produce biased results in combination
with the single-cluster update. In this case, we cannot check
Eq.~(\ref{SUPERTEST}) directly, as the deviations in the energy and
specific-heat for SC-{\bf PR} are not large compared to the errors.
Nevertheless, we can compare the bias for the SC-{\bf R250} in
the $16^3$ and $24^3$ lattices (see table \ref{SESGO3D}), 
which is a test of the $L$ dependence of the linear coefficient
in Eq. (\ref{SUPERTEST}). From Finite-Size Scaling theory, we can
estimate that
\begin{equation}
\frac{(\Delta C_v/\Delta E)_{L=24}}{(\Delta C_v/\Delta E)_{L=16}} 
\approx \left(\frac{24}{16}\right)^{1/\nu}\approx 1.9 \, ...
\end{equation}
where $\nu\approx 0.63$ is the critical exponent for the
correlation-length.  From table \ref{SESGO3D}, the above quotient can
be estimated to be $1.7(5)$, which is certainly compatible with our
prediction, but the error is so big that this is not a compelling
evidence for Eq. (\ref{SUPERTEST}).  Now, if we assume again
Eq.~(\ref{SUPERTEST}), we obtain for the $16^3$ lattice $\Delta E
\approx -5 \Delta {\mathrm A}_1$ and $\Delta C_v \approx 194 \Delta
{\mathrm A}_1$ From these relations, and from the estimate of $\Delta
{\mathrm A}_1^{{\mathrm {SC-PR}}}$ we obtain for the
bias (the statistical errors in fig. \ref{163D} being $\sigma_E$ and 
$\sigma_C$)
\begin{equation}
{\begin{array}{ccccccc}
\Delta E^{{\mathrm {SC-PR}}}&\approx&0.00015\, &,&
\sigma_E&=&0.00019\, ,\\
\Delta C_v ^{{\mathrm {SC-PR}}}&\approx&0.007\, &,&
\sigma_C&=&0.005\, .
\end{array}}
\end{equation}
Thus, it is not surprising that the bias does not show up in
Fig.~\ref{163D}.
For the $24^3$ lattice we lack an accurate measure of the bias
for ${\mathrm A}_1$, and so we cannot obtain a bias estimate.

As a final remark, notice that the sign of the bias seems to be
independent of the lattice size and the space dimension for {\bf
R250}.  This seems to be consistent with the simple (unidimensional)
model proposed in ref. \cite{BLOTE}.  However in the {\bf PR} case,
the (much smaller) bias changes sign when going from 2 to 3
dimensions. This suggests that the reason for bias is more involved in
this case.

\section{Conclusions}
In this work, we have shown that some Schwinger-Dyson identities,
Eqs. (\ref{UNOGLOBAL},\ref{ENERGIA}), are sensitive test of PRNG
induced bias. Most important, they can be used when no exact solution
is at hand.  We have provided some empirical evidence for a simple
relation between the bias induced in the different observables 
(our Eq.~(\ref{SUPERTEST})). This relation is obtained under the
assumption that the main effect of the bias is to produce a shift
on the coupling. It might be possible to justify this in terms of
relevant and irrelevant operators, in the framework
of the Renormalization Group. Furthermore, this suggests that an investigation
along the lines of Ref.~\cite{TONY} could be useful to establish
which new couplings are generated by the PRNG-induced bias.
If this relation could be established, the SD equation test would
provide an estimate on the maximum {\em safe} accuracy that one
can get for any observable, with the given PRNG.

In three dimensions, where there is no exact solution at hand, the
Schwinger-Dyson Equations test has shown that the single-cluster update with
the {\bf R250} and {\bf PR} PRNG's produces biased results, without
resource to seven more simulations.  It should be noticed that the
measure of the Schwinger-Dyson equations comes almost for free, as the
number of possible exponential factors is finite, and the local energy
should be measured anyway. Disk storage is not a shortcoming either,
because no reweighting~\cite{REWEIGHT} is to be done, and the
calculation can be made ``on the fly''. They are also extremely
helpful for code debugging. So, we believe Schwinger-Dyson equations
to be very useful tools, which can be easily measured in almost every
circumstances.

\section{Acknowledgments}

We acknowledge interesting discussions with L. A. Fern\'andez,
J. J. Ruiz-Lorenzo and A. Mu\~noz-Sudupe.

The computations have been carried out using the RTNN machines at
Universidad de Zaragoza and Universidad Complutense de Madrid. We
acknowledge CICyT for partial financial support (AEN97-1708 and
AEN97-1693).

\begin{table}
\caption{The bias for $E, C_v, {\mathrm A}_1, {\mathrm A}_2$, for the
$16^2$ lattice simulated with the SC-{\bf R250}, SC-{\bf PR}
combinations. To be able of measuring the bias in the SC-{\bf PR}
combination we have needed a much longer simulation (see text). The
constancy of the ratios is a check for Eq.~(\ref{SUPERTEST}).}
\begin{tabular}{lllll}
& $\Delta E$ 
& $\Delta C_v$ 
& $\Delta {\mathrm A}_1$
& $\Delta {\mathrm A}_2$  \\\hline
SC-{\bf R250} & -0.00235(11) & 0.0599(14) & 0.00148(19) & -0.0029(4)\\
SC-{\bf PR}   & -0.00057(2)  & 0.0115(2)  & 0.00024(4)  & -0.00047(8)\\\hline
Ratio         &  0.242(14)   & 0.192(5)   & 0.16(3)     &  0.16(4) \\
\end{tabular}
\label{SESGO2D}
\end{table}

\begin{table}
\caption{Bias for $E, C_v, {\mathrm A}_1, {\mathrm A}_2$, for the
$16^3$ and $24^3$ lattices simulated with the SC-{\bf R250}
combination.The ``correct'' value has been taken from the averaged
estimate of Figs. \ref{163D} and \ref{243D}.  The 
ratios test the lattice-size dependence of the 
coefficients in Eq.~(\ref{SUPERTEST}).}
\begin{tabular}{lllll}
$L$
& $\Delta E$ 
& $\Delta C_v$ 
& $\Delta {\mathrm A}_1$
& $\Delta {\mathrm A}_2$  \\\hline
$16^3$ & -0.00124(19) & 0.051(4) & 0.00026(5) & -0.00082(16)\\
$24^3$ & -0.00066(13) & 0.046(5) & 0.00014(3) & -0.00044(11)\\\hline
Ratio  & 0.53(13)    & 0.90(12) & 0.54(16)   &  0.54(4) \\
\end{tabular}
\label{SESGO3D}
\end{table}

\newpage

\newpage

\newpage

\begin{thebibliography}{99}


\bibitem{PERC} H. G. Ballesteros, L. A. Fern\'andez,
V. Mart\'{\i}n-Mayor, A. Mu\~noz Sudupe, G.~Parisi and
J. J. Ruiz-Lorenzo, {\sl Phys. Lett.} {\bf B400}, 346 (1997).

\bibitem{VATTU} I. Vattulainen, T. Ala-Nissila and K. Kankaala, {\sl
Phys. Rev. Lett.} {\bf 73}, 2513 (1994).

\bibitem{R250} G. A. Marsaglia in {\sl Computational Science and
Statitics: The interface}, ed. L. Balliard (Elsevier, Amsterdam,
1985); S. W. Golomb, {\sl Shift Register Sequences } (Holden-Day, San
Francisco, 1967).

\bibitem{PARISIRAPUANO} G. Parisi and F. Rapuano, {\sl Phys. Lett.}
{\bf B157}, 301 (1985)

\bibitem{DPLANDAU} A. M. Ferrenberg, D.P. Landau and Y. J. Wong, {\sl
Phys. Rev. Lett.} {\bf 69}, 3382 (1992).

\bibitem{WOLFF} U.~Wolff, {\sl Phys. Rev. Lett.} {\bf 62}, 3834
(1989).


\bibitem{BLUMECAPEL} F. Schmid and N.B. Wilding, {\sl
Int. J. Mod. Phys.} {\bf C6}, 781 (1995).

\bibitem{CODDINGTON} P.D. Coddington, cond-mat/930917.

\bibitem{FISHER} A. E. Ferdinand and M. E. Fisher, {\sl Phys. Rev.}
{\bf 185}, 832 (1969).

\bibitem{KOGUT} M. Baig, H. Fort, J. Kogut and S. Kim, {\sl
Phys. Rev.} {\bf D51}, 5216 (1995).


\bibitem{BLOTE} L. N. Shchur and H.W.J. Bl\"ote, {\sl Phys. Rev.} {\bf
E55}, 4905 (1995).

\bibitem{TONY} A. Gonzalez-Arroyo and M. Okawa, {\sl Phys. Rev. Lett.}
{\bf 58}, 2165 (1987).

\bibitem{METROPOLIS} N. Metropolis, A. W. Rosenbluth,
M. N. Rosenbluth, A. H. Teller and E. Teller, {\sl J. Chem. Phys.}
{\bf 21}, 1087 (1953).

\bibitem{CLUSTER} R. H. Swendsen and J. S. Wang, {\sl
Phys. Rev. Lett.} {\bf 58}, 86 (1987).

\bibitem{TALAPOV} A. L. Talapov and H. W. Bl\"ote, {\sl J. Phys. }
{\bf A 29} 5727 (1996).

\bibitem{REWEIGHT} M. Falcioni, E. Marinari, M. L. Paciello, G. Parisi
and B. Taglienti, {\sl Phys. Lett.} {\bf 108} 331 (1982) ;
A. M. Ferrenberg and R. H. Swendsen, {\sl Phys. Rev. Lett.} {\bf 61}
2635 (1988).


\end{thebibliography}
\end{document}